# COMET: A Cross-Layer Optimized Optical Phase Change Main Memory Architecture

Febin Sunny*, Amin Shafiee*, Benoit Charbonnier+, Mahdi Nikdast*, and Sudeep Pasricha*
*Colorado State University, USA,
+Université Grenoble Alpes, France
febin.sunny@colostate.edu, amin.shafiee@colostate.edu, benoit.charbonnier@cea.fr,
mahdi.nikdast@colostate.edu, sudeep@colostate.edu

*Abstract*— Traditional DRAM-based main memory systems face several challenges with memory refresh overhead, high latency, and low throughput as the industry moves towards smaller DRAM cells. These issues have been exacerbated by the emergence of data-intensive applications in recent years. Memories based on phase change materials (PCMs) offer promising solutions to these challenges. PCMs store data in the material's phase, which can shift between amorphous and crystalline states when external thermal energy is supplied. This is often achieved using electrical pulses. Alternatively, using laser pulses and integration with silicon photonics offers a unique opportunity to realize high-bandwidth and low-latency photonic memories. Such a memory system may in turn open the possibility of realizing fully photonic computing systems. But to realize photonic memories, several challenges that are unique to the photonic domain such as crosstalk, optical loss management, and laser power overhead have to be addressed. In this work, we present *COMET*, the first cross-layer optimized optical main memory architecture that uses PCMs. In architecting *COMET*, we explore how to use silicon photonics and PCMs together to design a large-scale main memory system while addressing associated challenges. We explore challenges and propose solutions at the PCM cell, photonic memory circuit, and memory architecture levels. Based on our evaluations, *COMET* offers 7.1× better bandwidth, 15.1× lower EPB, and 3× lower latencies than the best-known prior work on photonic main memory architecture design.

## I. INTRODUCTION

Over the past several decades, the emergence of big data and machine learning workloads has given rise to massive data-driven applications. These applications, which include large language models [1], intrusion detection systems [2], and graph processing frameworks [3], [4], consume and generate data at unprecedented rates, requiring data storage in the order of terabytes (*TB*) and memory bandwidths in the order of *TB/s*. Conventional electronic memory technologies such as dynamic random-access memory (DRAM) are struggling to keep up with such demands for increasingly higher bandwidth [6] and energy efficiency [7]. Additionally, DRAM technology also faces challenges associated with scaling towards the 10-nm technology node. Current DRAM nodes, such as Micron's 1α and 1β, are fabricated at 12–14 nm. At lower node scales, it has been shown that the DRAM cell's charge retention diminishes, cell structural integrity deteriorates, and delay and power penalties associated with bit lines increase dramatically [6]. While 3D-stacking technologies and through silicon vias have enabled high bandwidth memory (HBM), the increasing demand for capacity, throughput, and energy efficiency warrants the exploration of new main memory technologies.

Non-volatile memories (NVMs) address the data retention challenges in DRAMs and can help avoid the need for refreshes and associated latency concerns. But NVM candidates based on ferroelectric (FRAM) [8] and resistive metal oxide (RRAM) [9] technologies generally suffer from reliability and write endurance issues. To achieve higher reliability while retaining the advantages that NVMs offer, NVMs based on phase change materials (PCMs) can be considered [10]-[12]. PCM cells show higher energy efficiency, bit densities, and bandwidth than other NVM cell types [13], [14]. PCMs can transition between two material states: amorphous and crystalline. These states offer high resistance contrast between them and hence can be used to store data as resistance levels. In electrically controlled PCM (EPCM) cells, the phase transitions are brought about by using current pulses. The state transition between amorphous and crystalline can be controlled to achieve different levels of crystallization of the PCM to achieve multi-level cells (MLCs) as well. But relying on PCM resistance as a way to represent data has caveats. The resistance levels achieved in PCMs have a non-linear dependence on the write voltage [15]. This makes achieving an intermediate state, between the fully amorphous and fully crystalline states, challenging. Furthermore, the written resistance level can also face resistance drift, limiting electrical PCM bit density to just a few bits per cell [16].

One solution to these limitations with EPCM cells is to utilize optically controlled PCM (OPCM) cells, where the PCM is deposited on top of a photonic (e.g., silicon-on-insulator (SOI)) waveguide. In such OPCM cells, state transitions can be achieved by using laser pulses. The power delivered by the laser pulses can heat up the material, enabling state changes between amorphous and crystalline states. The refractive index contrast between the states affects the optical transmission of the cell, enabling storing and reading out data optically. If the PCM candidate selected has a sufficiently high contrast between the two states, intermediate states between amorphous and crystalline states can be used to create PCM-based MLCs. Moreover, an optically controlled PCM memory comes with the added advantage of being able to leverage high bandwidth silicon photonic links for data transfer. Given the emergence of optical computing [18], an optical memory can also enable high-speed and energy-efficient fully photonic computing systems with minimal electro-optic conversions.

In this paper, we present the design of the first cross-layer optimized optical PCM-based main memory architecture, named *COMET*. The proposed main memory system is characterized by high bit density per cell, lower energy consumption, and high memory throughput and bandwidth compared to the state-of-the-art. Our novel contributions in this work are:

- We comprehensively explore different PCM candidates with the goal of selecting the most efficient PCM for optical memory applications;
- We design a low-loss and energy-efficient silicon photonic PCM-based multi-level memory cell as a basic building block;
- We design and optimize an all-optical, loss-aware silicon photonic PCM-based photonic main memory architecture; and
- We present detailed comparison of our designed photonic main

memory architecture against state-of-the-art electronic and photonic main memory architectures.

## II. BACKGROUND AND RELATED WORK

In this section, we discuss the fundamentals of PCMs and PCM-based optical memories.

### A. PCM: Fundamentals and Properties

PCMs are capable of changing phase from amorphous to crystalline, and vice versa, based on the thermal energy supplied to the material. The thermal energy supplied should be sufficient to change the temperature across the bulk of the material. This change in temperature should match the melting temperature ($T_l$; for phase change to amorphous state) or the crystallization temperature ($T_g$; for phase change to crystalline state). Amorphization is the more power-hungry process as $T_l > T_g$. Additionally, PCMs can be set to an intermediate state if the provided thermal energy converts only part of the material to either amorphous or crystalline state [12]. The energy necessary to achieve these state transitions can be delivered to the PCM electrically, thermally, or optically. Microheaters can be used for applying thermal power directly, while PN junctions can be used to supply heat electrically. To trigger phase transitions optically, a laser pulse is required. The specific power and duration at which the laser pulse delivers thermal energy depend on the material and the energy required by that material to achieve state transition. Three widely considered phase-change materials in prior work include $Ge_2Sb_2Te_5$ (GST), $Ge_2Sb_2Se_4Te$ (GSST), and $Sb_2Se_3$ [12].

The change in PCM phase brings with it a change in the electrical and optical properties of the material. PCM's states have different electrical resistances. Typically, the high-resistance amorphous state is used to represent a binary 0, and the low-resistance crystalline state is used to represent a binary 1. This non-volatile change in resistance allows the PCM cell to be paired with an access transistor to form a 1T-1R EPCM memory, as described in many prior works (e.g., [31], [40]-[42]). But as discussed earlier, EPCM memories face many challenges, such as asymmetric and high write latencies [19], non-linear response to write voltage, and resistance drift.

Optical PCM-based (OPCM) memories depend on the change in the refractive index of the material phases. The change in *refractive index* changes the optical transmission across the cell, which allows data storage and readout. To implement such an OPCM memory effectively, understanding the optical properties of the PCM material is important. For PCM candidates, high refractive index contrast and hence contrast in optical transmission between the phases is essential. Having higher refractive index contrast between the amorphous and crystalline state enables better tolerance to optical signal losses and noise which can otherwise cause readout errors. This high contrast also allows multiple intermediate levels of phase transition to be achieved without them being susceptible to these same losses. In this regard, achieving high refractive index contrast serves the same purpose as achieving high resistance contrast between phases in EPCM memories, and leads to better signal-to-noise ratio (SNR) at the readout.

A higher *extinction coefficient* between states is also an important requirement in an OPCM cell. Extinction coefficient in photonics is a measure of the optical power dropped from an optical signal as it traverses a material. Higher extinction coefficient indicates that the material extracts more power from the signal. The benefit of this metric depends on the application. From the perspective of propagation (e.g., in optical interconnects), high extinction coefficient is not preferred as this results in higher losses and power consumption, but for a filter (e.g., in optical switches), high extinction coefficient means that the device will be able to filter out as much of the optical signal as possible. In OPCM memory applications, a higher extinction coefficient in memory cells is beneficial. More efficient laser power absorption due to a high extinction coefficient in the crystalline state allows for energy-efficient transition to the amorphous state, and vice-versa.

### B. OPCM Memory

Given a specific phase change material, it is essential to design an efficient memory cell that can grant access to the material for reads and writes. There are several ways in which the OPCM cell design has been approached in the literature, as discussed next.

The work in [20] proposed a simple crossbar-based cell design (Fig. 1(a)) where the OPCM material is placed on top of waveguide crossings. The work proposed a main memory architecture called *COSMOS* using this OPCM cell design. Access to the cell in this architecture is provided through row and column access wavelength signals. These signals have to be present simultaneously to ensure write operations. The proposed architecture employed a subtractive read approach where the entire subarray is read, followed by a reset signal to the row that needs to be read, erasing the contents, and finally, the subarray is read again. The two read values are then subtracted at the memory controller (MC) to obtain the intended row values. This approach paired with the 4 bits/cell assumption ensures a high bit density for such an architecture.

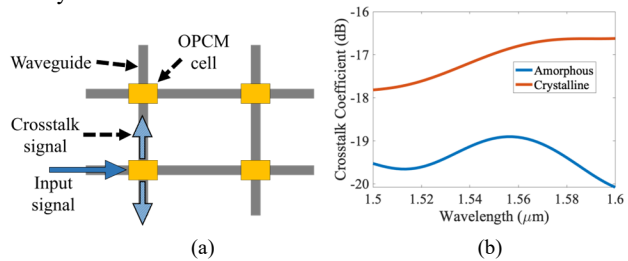

Fig. 1. (a) OPCM array structure from [20]; (b) Crosstalk experienced in (a).

However, the cell design in the *COSMOS* architecture from [20] causes the rows to be susceptible to crosstalk from the write operations on the adjacent rows. The crosstalk signal (Fig. 1(a)), although small, can trigger changes in the OPCM cell due to the thermo-optic effect [21]. The energy from the write pulses can cause temperature changes and hence refractive index changes in the adjacent cells, causing severe data corruption. This effect is exacerbated when multiple bits are stored per cell, as a small change in the OPCM's refractive index can considerably impact the data stored. *COSMOS* uses GST cells designed in [17] which require up to 750 pJ to operate. But *COSMOS* assumes a 135 pJ operational energy which is insufficient for GST cell operation. Even with a 750 pJ energy delivered at the crossbar, the thermo-optic effect due to the ~ (–18 dB) crosstalk (see Fig. 1(b)) can introduce 12.6 pJ energy to the adjacent cells in the *COSMOS* architecture. This extraneous energy can trigger an 8% change in a neighboring OPCM cell's refractive index, which can easily alter data stored in a cell with 16 programmable refractive-index levels (i.e., 4 bits/cell) [21] with the < 8% contrast between levels assumed in *COSMOS*. Without corrective measures after every write operation, data stored in the *COSMOS* architecture can get severely corrupted as shown in Fig. 2. This is without considering the fact that a contiguous array of GST cells that read out data through other cells in the row will have to account for the optical losses as the signal passes through subsequent GST cells. Without accounting for these variable losses, as different GST cells will store different data and hence create different losses, the readout from the top row of the array may not reach the controller for detection. These losses can range from 0.24 dB for the amorphous state cells to as much as 21.8 dB for cells in the crystalline state. This further renders the proposed read and write approaches in [20] prone to severe error.

The issues highlighted above make the crossbar-based cell design from [20], although attractive from the bit density perspective, unusable from a memory reliability perspective. To ensure proper data retention and readout, the memory cells have to be isolated and access control mechanisms need to be in place. This can be achieved by using microring resonators (MRs) as access control mechanisms for these cells, as described in [22], [23]. These works proposed using thermo-optic tuning to regulate MR operation and grant photonic access to the cell. However, thermal tuning [24] to ensure that the MR is in-resonance (access granted) or off-resonance (no access to cell) is a slow process with microsecond ($\mu s$)-scale latencies. Although this is still much faster than the refresh mechanisms that have millisecond($ms$)-scale latencies in DRAMs, the tuning has to be performed every time a cell is accessed and hence it will severely increase the latency and reduce achievable bandwidth of the OPCM memory. In our design (described in the next section), we propose using electro-optic tuning [25], where the resonance of the MR is controlled using carrier injection through a PN junction at nanosecond($ns$)-scale latencies. Note that this imposes increased optical losses which must be accounted for in our design considerations, which is discussed in the next section.

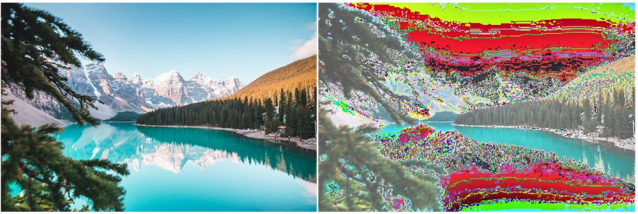

Fig. 2. Data corruption in crossbar-based OPCM memory from [20] due to crosstalk; (*left*) original image; (*right*) image after 4 writes to adjoining rows.

Once the cell design is finalized, multiple cells can be tiled to form subarray and bank structures to realize a photonic main memory architecture. We discuss our material selection, cell design, and architecture-level design for *COMET* in the next section.

## III. *COMET* OPCM-BASED MAIN MEMORY DESIGN

In this section, we describe the components of our proposed *COMET* OPCM-based main memory architecture. Our memory architecture ensures reliable data writes and reads while achieving high data throughput. The memory cells have access control mechanisms that isolate the cells from each other and ensure crosstalk elimination, to prevent data corruption. We also adopt a combination of mode-division multiplexing (MDM) and wavelength-division multiplexing (WDM) to enable parallel accesses for reads and writes. This access mechanism allows *COMET* to be designed as a hierarchical multi-banked design, rather than a simple array of memory cells.

### A. Phase Change Material Selection

The refractive index contrast and extinction coefficient are two of the most critical design parameters to consider when determining the most suitable material for OPCM-based main memory applications, as discussed in Section II.A. The refractive index contrast and extinction coefficient of three well-known PCM candidates $Ge_2Sb_2Te_5$ (GST), $Ge_2Sb_2Se_4Te$ (GSST), and $Sb_2Se_3$ can be modeled using the Lorenz model [27]. We modeled and analyzed the two key design parameters for the PCM candidates, with results shown in Fig. 3. It can be observed that for C-band (1530–1565 nm), GST exhibits the highest refractive index contrast (difference between blue and yellow lines, and difference between red and purple lines) and high extinction coefficient between amorphous and crystalline states. This makes GST the most suitable candidate for OPCM-based memory cells. Thus, we consider GST for our cell design. Prior works have also demonstrated that it is possible to store more than 34 unique states in GST-based OPCM memory cells [17], thus enabling up to 5 bits/cell capacity in OPCM memories.

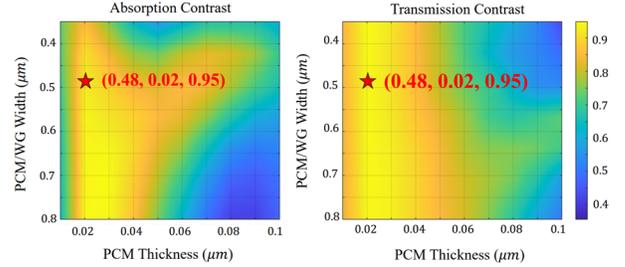

Fig. 4. Optical absorption contrast and optical transmission contrast of GST cell for different cell geometry (width and thickness) in the *COMET* architecture. The stars represent the geometric configuration selected with values for (width, thickness, absorption or transmission contrast ratio), based on our analysis.

### B. OPCM Memory Cell Design

In our OPCM cell design, we consider GST deposited on a silicon-on-insulator (SOI) strip waveguide. As it was shown in Fig. 3, the extinction coefficient of the GST in the crystalline state is much higher than its amorphous state. This leads to negligible optical transmission due to high absorption of the electric field in the PCM. The optical absorption contrast and optical transmission contrast between fully crystalline and fully amorphous state for OPCM memory cells of different geometries and materials are shown in Fig. 4. Note that the optical transmission contrast is not only a function of optical absorption in the cells but also partially originates from the optical-refractive-index mismatch between the PCM and SOI waveguide due to high refractive-index contrast between silicon and GST. To avoid optical-refractive-index mismatch when designing GST-based OPCM memory cells, it is important to select a design where both optical transmission contrast and optical absorption contrast are maximized. Doing so ensures that the optical transmission contrast stems from the optical power absorption which can be controlled by crystallization of the GST.

For a 2 µm long GST cell considered in our study (Fig. 4), optical transmission and absorption are at 95% when the thickness of the cell is 20 nm. Note that the impact of PCM waveguide (WG) width on optical transmission and absorption is negligible. The SOI waveguide has a width of 480 nm (to ensure the single mode transmission of the light) and a thickness of 220 nm, where the GST deposited on it has the same width, but a thickness of 20 nm (Fig. 5(a)). We have designed the GST cell with a small thickness, as a higher thickness makes heat transfer over the volume of the cell slower, hence leading to higher write and reset latencies. We also opted for a silicon (Si) waveguide instead of silicon nitride (SiN) waveguide, as Si offers higher transmission contrast between crystalline and amorphous states of the cell. Si also offers higher mode confinement over SiN waveguides, leading to lower propagation losses. Overall, using Si platforms can lead to a more compact GST cell footprint, lower amorphization time,

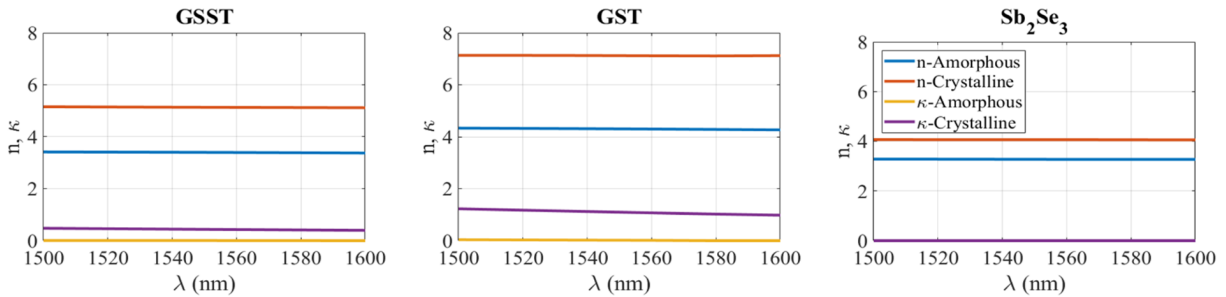

Fig. 3. Comparison of the refractive index ($n$) and extinction coefficient ($\kappa$) between GSST, GST, and $Sb_2Se_3$ in the optical C-band range.

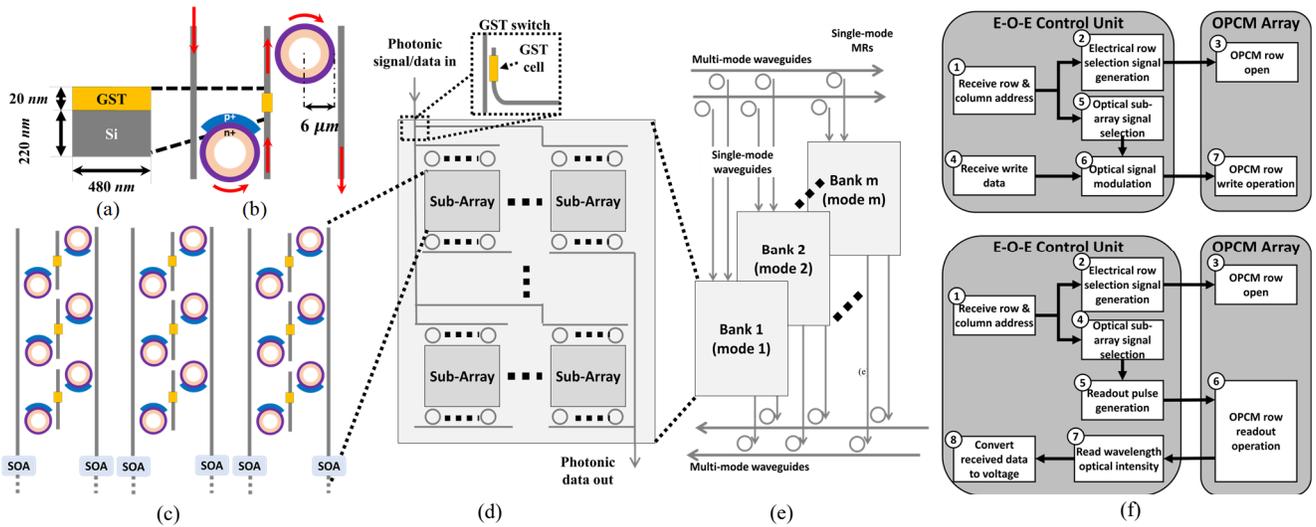

Fig. 5. (a) GST cell designed and simulated for this work. (b) Our proposed memory cell with MR-based access control. (c) OPCM memory array with intra-subarray semiconductor optical amplifiers (SOAs). (d) Single bank with multiple subarrays, with inset showing GST based signal switch. (e) Overall multi-bank architecture of *COMET*. (f) Steps during read (*top*) and write (*bottom*) operations in *COMET* architecture.

and higher transmission contrast between amorphous and crystalline states [12].

To obtain the optical transmission characteristics of the GST cell we designed, we used FDTD simulations from Ansys Lumerical FDTD [26]. Since we are interested in OPCM MLCs, we obtain the refractive index profile of the intermediate states of phase transition using the Lorenz model [27]. The modeled refractive index profile was imported to the FDTD simulator. From these FDTD simulations, we were able to obtain the optical transmission levels of the intermediate states.

To obtain the energy and latency of transitions for intermediate states, Ansys/Lumerical HEAT [26] was used to solve transient unsteady-state heat transfer equations to capture the time-dependent temperature distribution over the OPCM's cell volume. In this simulation, a local uniform heat source was defined in the Si waveguide to mimic the power of the optical mode propagating in the waveguide. The regions of the GST cell which have a temperature between $T_l$ and $T_g$ have a crystalline structure, whereas the regions with temperatures above $T_l$ exist in an amorphous state because of the melt and quench mechanism [17]. We have considered two case studies, for two possible programming modes for an OPCM multi-level cell: (1) when the deposited state is crystalline, hence the reset state will be crystalline, and (2) when the deposited state is amorphous, hence the reset state is amorphous. Using the aforementioned methodology, a 4-bit OPCM GST cell with 16 distinctive and equally spaced transmission levels (with 6% spacing between transmission levels) was simulated. From these simulations, latency of transition, crystalline fraction in the transition level, and the optical transmission of the transition level were obtained, as shown in Fig. 6. For case study (1), the reset pulse required 880 pJ of energy and for (2), the reset pulse required 280 pJ of energy.

As our architecture considers WDM-based data transfers, the performance of the GST cell, when exposed to different wavelengths, needs to be analyzed. Results for the C-band (1530–1565 nm; plots not shown for brevity) showed a linear decrease of loss from 0.073 dB/mm to 0.067 dB/mm for the wavelength range between 1530 nm to 1565 nm. In addition, the maximum wavelength-dependent transmission contrast between crystalline and amorphous states was calculated to be 1.4%. These results indicate that our GST cells can perform with low loss and low variation in transmission levels across the C-band.

After finalizing the GST cell design, we designed the OPCM memory cell which integrates the GST cell and regulates signal access to the cell. This access control provides cell isolation between adjacent GST cells and is necessary to avoid optical crosstalk and associated thermo-optic-effect-based data corruption, as discussed in Section II.B.

To ensure GST cell isolation, our memory cell has MR-based access control as shown in Fig. 5(b). MRs are switched in and out of resonance to enable and disable the signal access to the GST cell. We use MR designs from [36] with 6 μm radius, for low loss and efficient coupling during EO tuning to allow as much of the read/write signal to reach the GST cell from the Si waveguides. The signal enters the cell through the left waveguide and follows the path shown using red arrows in Fig. 5(b) and exits through the rightmost waveguide. As discussed in Section II.B, we have opted for electro-optic tuning for tuning the MRs in and out of resonance for the 2 ns access latencies this access mechanism provides [36].

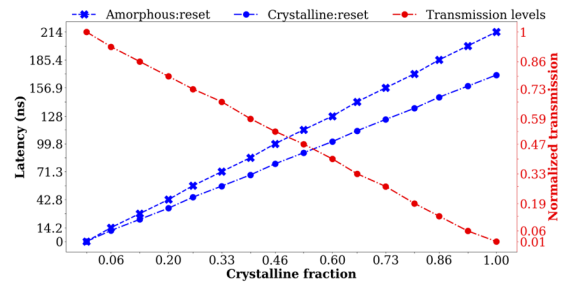

Fig. 6. Latency and optical transmission for 16 crystalline-fraction levels representing the intermediate states in our designed GST cell.

### C. COMET Memory Bank Architecture Design

The OPCM memory cell can be tiled to form an array, as shown in Fig. 5(c). The array is comprised of columns of the designed OPCM memory cell, with each column assigned a single wavelength for accessing the GST cell. With this arrangement, the GST cells stay isolated from each other. The row access of the array is provided through simultaneous electrical control of the PN junctions of the MRs in the row, shifting them into resonance and, in doing so, allowing the wavelengths access to the GST cell.

The memory bank thus formed has $N_r \times N_c$ OPCM cells, with a total capacity of $N_r \times N_c \times b$, where $b$ is the bit capacity of the OPCM MLC. This bank has to be further divided into subarrays to enable energy-efficient access. A bank will contain $S$ subarrays, each containing $M_r \times M_c$ cells, such that $N_r = S_r \times M_r$ and $N_c = S_c \times M_c$. When a subarray has to be accessed, $M_c$ wavelengths need to be modulated, and $2 \times M_c$ MRs per bank have to be tuned to be in resonance. The memory bank requires $N_c$ wavelengths to operate. We consider an off-chip laser to provide the $N_c$ wavelengths in this work. Consequently, $N_c$ MRs are needed to allow access to the columns, and another $N_c$ MRs are needed to allow readout from the columns.

To access such a subarray system, optical splitters and couplers are required, which essentially multiply the laser power needed. In *COMET*, we utilize electrically controlled GST-based waveguide switching [39] to allow efficient access to our subarrays. The GST cell is inserted at the waveguide coupler and is switched from crystalline (no coupling) to amorphous state (coupling) to allow access to the subarray (Fig. 5(d)). This adds a loss of 0.2 dB (for amorphous GST) when the wavelengths are coupled, and a GST switching time of 100 ns, but significantly reduces overall laser power requirement. The wavelength modulation and the MR tuning signals are generated at the electrical control unit, which interfaces the OPCM memory banks to the memory controller and processor. As general-purpose computing is still only feasible using electrical circuits, this electrical interface is necessary.

Note that in an $M_r \times M_c$ subarray, to preserve the integrity of the data being read and to ensure that the data signals reach the electrical interface, several design choices must be made. Depending on $b$, the data being read from the cells can only suffer so much loss before the transmission level becomes similar to the expected transmission level for other data. For example: For $b = 2$, the transmitted signal can suffer up to 25% or 1.2 dB of losses before a readout of '10' becomes the same as the readout for '01'. For $b = 4$, this tolerance becomes even lower, with the signal only being able to suffer less than 6% losses or 0.26 dB before the readout becomes erroneous.

We integrate semiconductor optical amplifier (SOA) based gain tuning within and outside the banks and subarrays to address data integrity issues (see Fig. 5(c)). At the electrical interface level, there needs to be row-wise loss-aware signal amplification so that the row-wise losses can be accounted for to preserve data integrity. Since these losses and the expected tolerances will be fixed once the design is finalized, for gain tuning at run-time, a look-up-table (LUT) based required gain storage is utilized. Based on the row address, the LUT can provide the gain necessary to tune the modulated wavelength, using SOAs. However, this gain may not be sufficient to survive the losses within the $M$ rows, with the losses from the EO-tuned active MRs. This necessitates SOAs within the subarray at regular intervals to boost the readout signal. These intra-subarray SOAs need only provide the same amount of power as the input laser signal to the bank (1 mW for crystalline reset programming and 5 mW for the amorphous reset programming). These SOAs play an important role in compensating for the EO-tuned MR through losses, for both read and write signals. We consider the power and latency overheads of the SOA and LUTs in our analysis presented in Section IV.

*COMET* is architected as a multi-bank OPCM memory (see Fig. 5(e)), enabled through the combination of mode-division multiplexing (MDM) and wavelength-division multiplexing (WDM). We enable parallel access across banks, and the interleaved cache lines using MDM. This requires designing silicon photonic links with an MDM degree of $B$ to enable this access. There are several considerations to be made in selecting the MDM degree. The fundamental mode is the mode that gives the highest confinement of the electrical field in the waveguide and hence the lowest loss. As the number of modes increases, the confinement of the higher modes decreases, since their effective index decreases. This causes the higher modes to suffer from higher losses. As higher order modes are excited, the field will become leakier with higher losses. In addition, to support higher order modes, the waveguide width of the links and devices also needs to increase. Prior works have shown an MDM degree of 4 is achievable on chip, without notable losses or area overhead [28]. So, for our architecture, we set the MDM degree (and hence $B$) to 4, to minimize overhead.

*D. Read and Write Operations in COMET*

For the read operation, we perform the steps as depicted in Fig. 5(f), top. The isolated nature of our memory cells makes the read operation straightforward, with row access through EO tuning and column access through sending the readout pulses to the subarray. The row ID is obtained from the address and the EO tuning signals are sent to the MRs in that row. Depending on the subarray ID mapped from the physical address, the corresponding wavelengths are gain-tuned and sent to the OPCM banks. Once the data is received at the electrical interface it is demodulated using an MR bank and the received data can be passed on to the processor for further operation.

Similarly, for the write operation (see Fig. 5(f), bottom), we follow a straightforward approach. The row ID is obtained from the address and row access is provided through tuning the MRs of the row into resonance. The column IDs are also obtained from the address and the corresponding wavelengths are gain-tuned and modulated to reflect the data to be written and sent to the OPCM subarray.

*E. COMET Power Consumption*

The *COMET* architecture can achieve a capacity of $(B \times N_r \times N_c \times b)$ bits. With the architecture design, the power required to achieve this capacity can be calculated as follows. With our SOA-based loss mitigation strategy, we assume $M_c = N_c$, which implies, $S_c = 1$. The subarrays can be arranged in an array of $\sqrt{S_r} \times \sqrt{S_r}$ for layout and addressing. The $M_r$ value would depend on the $b$ value to ensure the cache line readout across the $B$ banks. We consider intra-subarray SOAs which are able to provide a gain of 15.2 dB to their input signal [29]. Given that the EO-tuned MRs have a through loss of 0.33 dB, there needs to be an SOA array at every 46 rows. This necessitates a total SOA count of $\frac{B \times N_r \times N_c}{46}$. To minimize power overhead, we only enable the SOAs within the subarrays that are being accessed. Assuming 1.4 mW power consumption for 0 dBm i.e., 1 mW output [29], total SOA power consumption at any instance during *COMET* operation will be $\left(\frac{B \times M_r \times M_c}{46} \times 1.4\right)$ mW.

Apart from these power considerations, there is a need for considering power consumption of the photonic links. Our WDM-MDM link requires $N_c$ wavelengths to operate and requires $2 \times B \times N_c$ MRs to access the OPCM arrays. These MRs can be completely passive as they need not perform any switching operations. The laser power consumption can be calculated based on the various losses the signal will experience on its way to and from the OPCM arrays. These losses and associated power overhead are discussed in Section IV.

Finally, we need to consider the power consumption by the EO tuning mechanism within the OPCM arrays. Given that the MRs need to be tuned only within the row of the subarray being accessed, the tuning mechanism will contribute to $(B \times 2 \times M_c \times P_{EO})$ W of power. Here, $P_{EO}$ is power consumption for EO tuning a single MR. This is further discussed in Section IV.

*F. Address Mapping in COMET*

To access the data within the memory banks, we need to perform an address mapping to the cells in our architecture. *COMET* can consider cache lines of various sizes (e.g., 32, 64, and 128 bytes), to reflect popular last level cache line sizes, interleaved across the $B$ banks. The mapping process should perform the following mapping:

$$\{Channel_{ID}, Row_{ID}, Bank_{ID}, Column_{ID}\} \rightarrow$$
$$\{Channel_{ID}, Subarray_{ID}, Subarray_{ROW}, Bank_{ID}, Subarray_{COL}\} \quad (1)$$

For our architecture, as described in Section II.B, the channel and bank IDs can remain the same. $Row_{ID}$ must be mapped to $Subarray_{ID}$ and $Subarray_{ROW}$ and $Column_{ID}$ to $Subarray_{ID}$ and $Subarray_{COL}$. The values for these parameters can be calculated as follows:

$$ID_1 = int\left(\frac{Row_{ID}}{M_r}\right) \quad (2)$$

$$ID_2 = int\left(\frac{Column_{ID}}{M_c}\right) \quad (3)$$

$$Subarray_{ID} = ID_2 \times \sqrt{S_r} + ID_1 \quad (4)$$
$$Subarray_{ROW} = (Row_{ID} \% M_r) \quad (5)$$
$$Subarray_{COL} = (Column \% M_c) \quad (6)$$

## IV. EXPERIMENTS AND EVALUATION

In this section, we describe our simulation setup, various design parameters considered in the design of the *COMET* architecture, and our approach for comparison with other main memory architectures.

We use a modified version of NVMain 2.0 [30] a main memory simulator that we heavily modified to accommodate 4-bit/cell MLC operation, our photonic memory configurations, and the addressing scheme. Our experiments and evaluations are based on an 8 GB main memory chip capacity. Performance metrics encompass application latency, bandwidth, and energy-per-bit (EPB). We benchmark *COMET* against *COSMOS* [15], a prominent OPCM main memory architecture, EPCM-MM [24], a proposed EPCM main memory architecture, and 2D and 3D configurations of DDR3 and DDR4 DRAMs (labeled as 2D_DDR3, 3D_DDR3, 2D_DDR4, and 3D_DDR4). Memory traces from the SPEC benchmark suite [32] are utilized for architecture evaluation. The various parameters we have considered for power modeling of the *COMET* architecture are provided in Table I.

TABLE I: Optical loss and power parameters considered for *COMET* power modeling.

| Loss parameters | Values |
|---|---|
| Coupling loss | 1 dB [33] |
| MR drop loss | 0.5 dB [34] |
| MR through loss | 0.02 dB [35] |
| EO tuned MR drop loss | 1.6 dB [36] |
| EO tuned MR through loss | 0.33 dB [36] |
| Propagation loss | 0.1 dB/cm [37] |
| Bending loss | 0.01 dB/90° [38] |
| SOA gain | 20 dB |
| Laser wall plug efficiency | 20% |
| **Power parameters** | **Values** |
| EO tuning power ($P_{EO}$) | 4 $\mu$W/nm [25] |
| Max. power at GST cell | 1 mW |
| Intra-subarray SOA power | 1.4 mW [29] |

### A. Modeling COMET Architecture

As described in Section III.C, *COMET* has a capacity of $B \times N_r \times N_c \times b$ bits, with the array of $N_r \times N_c$ GST memory cells divided into subarrays of size $M_r \times M_c$ memory cells for enabling parallel and energy-efficient access. For bit density $b$, there are works which show the GST cell should be able to achieve up to 5 bits/cell [17]. However, considering data distribution across cells, it is practical to consider bit densities in the multiples of 2, which allows for an even distribution of data across cells in a row and equal loss and crosstalk considerations to be made across all columns.

But as discussed in Section III.C, at an architecture level, to allow high bit density per cell, several considerations must be made. To determine the optimal $b$ value in *COMET*, we analyze how power, latency, bandwidth, and EPB varies for $b = \{1, 2, 4\}$. For $b = 1$, our architecture for 8 GB would be $(4 \times 4096 \times 512 \times 1024 \times 1)$, reflecting ($B \times S_r \times M_r \times M_c \times b$). For $b = 2$, this would be $(4 \times 4096 \times 512 \times 512 \times 2)$. Finally, for $b = 4$, this configuration would be $(4 \times 4096 \times 512 \times 256 \times 4)$. We have opted to reduce $M_c (= N_c)$ over $N_r (= S_r \times M_r)$ as $b$ increases, as reduced $N_c$ results in the reduction of both WDM-degree requirement and significant intra-subarray SOA power reduction. Modifying $N_c$ also allows *COMET* to retain its cache line capacity and deliver the same bandwidth across the designs.

There is also a variable LUT size requirement at the electrical-optical interface as $b$ changes. The LUT size depends on the row frequency at which the SOA gain tuning is required. For $b = 1$, the loss tolerance is less than 50%, as there are only two levels stored per GST cell. The signal exiting the GST cell can suffer up to 3.01 dB of losses before it becomes error-prone. With MR through loss of 0.33 dB, the signal can pass 9 rows other than the source row, without error. For $M_r$ of 512, the LUT requires 52 entries. Given the intra-subarray gain tuning at every 46 rows, these entries can be repeating values. Hence making the entry requirement just 5 parameters, with the gain parameter selected as per the $ceil\left(\frac{rowID\%46}{10}\right)^{th}$ entry of the LUT. For $b = 2$, the loss tolerance is lower at 25% or 1.2 dB. Following the same process as for $b = 1$, the LUT requires 12 entries, with the gain parameter selected as per the $ceil\left(\frac{rowID\%46}{4}\right)^{th}$ entry of the LUT. For $b = 4$, the LUT requires 46 entries with the gain parameter selected as per the $(rowID\%46)^{th}$ entry of the LUT. Since the LUT is a part of the memory control interface, we do not consider this power consumption as part of the OPCM memory operation, even though it is negligible. The overall power requirements for the $b = \{1, 2, 4\}$ configurations can be calculated as per the discussion in Section III.E.

The resulting stacked power plots are shown in Fig. 7. Based on these analyses, we have chosen $b = 4$ to keep the power overhead relatively low, and hence the memory configuration of $(4 \times 4096 \times 512 \times 256 \times 4)$ is considered in our subsequent comparative analysis.

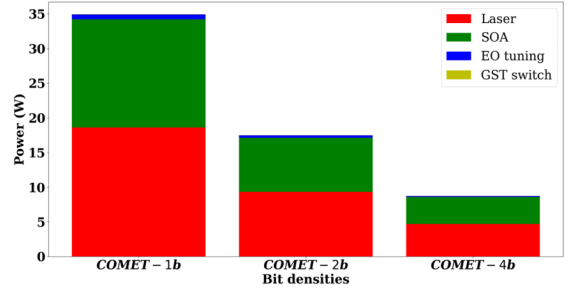

Fig. 7. Power stack-plots for COMET with bit density ($b$) of 1 (*COMET*-1$b$), 2 (*COMET*-2$b$), and 4 (*COMET*-4$b$).

### B. Modeling COSMOS Architecture

As *COSMOS* [20] is the only other photonic main memory architecture published in literature, we model it and compare our work against it. As discussed in Section II.B, there are several challenges with obtaining the correct readout data and ensuring correct writes with a crossbar OPCM architecture as presented in *COSMOS*. We model *COSMOS* and update its design assumptions to ensure correct readouts, which is essential to realize a practical main memory architecture.

TABLE II: Architectural details of photonic memory systems.

| COMET | 4 banks, 1 rank/channel, 1 device/rank<br>bus width = 256 bits, burst length = 4<br>max. write time = 170 ns, erase time = 210 ns, read time = 10 ns, data burst time = 1 ns, electrical interface delay= 105 ns |
|---|---|
| COSMOS | 8 banks, 1 rank/channel, 1 device/rank<br>bus width = 128 bits, burst length = 8<br>write time = 1.6 $\mu$s, erase time = 250 ns, read time = 25 ns, data burst time = 1 ns, electrical interface delay= 105 ns |

But before we address the crosstalk and data corruption issues in *COSMOS*, we must address the energy delivery assumptions. The GST cell design in *COSMOS* is taken from [17], which requires 5 mW laser pulses over 50 ns to 150 ns to deliver 250 pJ to 750 pJ in energy for phase transitions. *COSMOS* has retained the timing parameters from [17] but has reported the cells to require only 0.5 mW laser pulses. To ensure that the energy required for phase transitions is delivered to the memory cell, we have remodeled the timing constraints and assume 5 mW laser pulse power to ensure that the correct energy is delivered to the GST cells. We have opted to increase the timing parameters as increasing the power value would make the total power consumption entirely too high for an 8 GB main memory. The modified parameters for *COSMOS* (and also *COMET*) are provided in Table II.

Data corruption in *COSMOS* is a result of the thermo-optic effect and the losses that the optical signals experience as they traverse the crossbar. As discussed in Section II.B, the extraneous energy is sufficient to trigger an 8% shift in crystalline fraction, due to how the cells are exposed to each other. We decided to not re-architect *COSMOS* to enable cell isolation as this will depart too far from the design in [20]. We instead opt to change the intermediate level count to allow for higher tolerance to this 8% shift in crystalline fraction. This necessitates dropping *COSMOS* bit density *b* from 4 to 2. This results in a memory architecture where $(B \times N_r \times N_c \times b)$ is $(16 \times 16384 \times 16384 \times 2)$ and $S_r \times M_r = S_c \times M_c = 512 \times 32$. For this version of *COSMOS*, we make the generous assumption that the MDM losses are negligible as designing 16 degree MDM silicon photonic links without losses is extremely challenging.

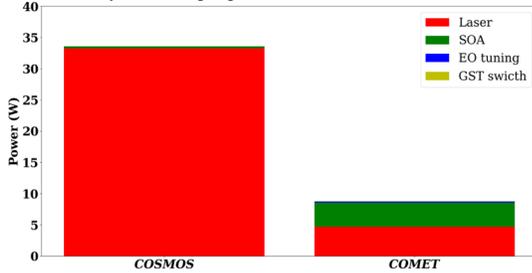

Fig. 8. Power stack-plot for *COSMOS* and *COMET* architectures.

Additionally, we have to consider the losses that the read/write signals face as they traverse the GST cell array. We select 4 asymmetric transmission levels (0.99, 0.90, 0.81, 0.72) separated by 9% transmission to avoid the thermo-optic corruption, to represent the 4 levels necessary to represent 2 bit data, to avoid the high losses at high crystalline fractions. Keeping the subarray size of $32 \times 32$ from COSMOS, the worst case loss of 1.4 dB (from transmission level 0.72) and the 15.2 dB gain for SOAs [29], this also requires 6 SOA arrays (when considering both row and column losses) per subarray. There also needs to be dedicated data signal in and out ports for these subarrays, to avoid further data corruption as the signal traverses the entirety of the 512×32 GST cells per row/column. We assume these are passive MR-based ports. To avoid excessive splitter loss and hence laser power consumption, we also assume a PCM cell based subarray row access control in *COSMOS* (which we proposed for *COMET* in this work), separating the subarray rows. Overall, this adds 0.2 dB PCM switch loss to laser power calculations, some additional SOA power consumption, and a 100 ns delay while accessing the subarray row to *COSMOS*, which is similar to the overhead for such access control in *COMET*. Fig. 8 shows the power stack comparison between *COSMOS* and *COMET*. The next subsection provides more detailed comparison results for latency, bandwidth, and EPB.

### C. Performance Evaluation

We compare *COMET* in terms of bandwidth and EPB against 2D_DDR3, 3D_DDR3, 2D_DDR4, 3D_DDR4, the EPCM-MM, and *COSMOS* modified from [20], as discussed in the previous subsection. The absence of refreshes along with higher bandwidth provided by the silicon photonic interconnects allow both *COSMOS* and *COMET* to outperform their electronic counterparts in terms of bandwidth, (Fig. 9(a)). *COMET* achieves 100.3×, 47.2×, 58.7×, 42.1×, 40.6×, and 5.1× higher bandwidth on average than 2D_DDR3, 3D_DDR3, 2D_DDR4, 3D_DDR4, EPCM-MM, and *COSMOS*, respectively.

In terms of EPB, it can be observed from Fig. 9(b) that the 3D and PCM electronic counterparts are able to outperform both fully photonic memory systems. This can be attributed to the fact that the entire power consumption depicted in Fig. 8 is utilized for orchestrating reads and writes for photonic memory. This is different from the case in an electronic memory where only a very small portion of the overall power is required to orchestrate a read/write. In order to achieve comparable EPB values, more intelligent read/write orchestrations with dynamic power control are necessary for photonic memories. From Fig. 8 it can be observed that laser power is a significant contributor to the overall power consumption of photonic memories. Enabling dynamic laser power management, such as that discussed in [43], could significantly improve photonic memory energy consumption. We leave the exploration of such techniques as future work. However, the high read/write bandwidth in comparison, enables *COMET* to outperform the 2D DRAM platforms. *COMET* is able to achieve 4.1×, 2.3×, 12.9× lower EPB than 2D_DDR3, 2D_DDR4, and *COSMOS*.

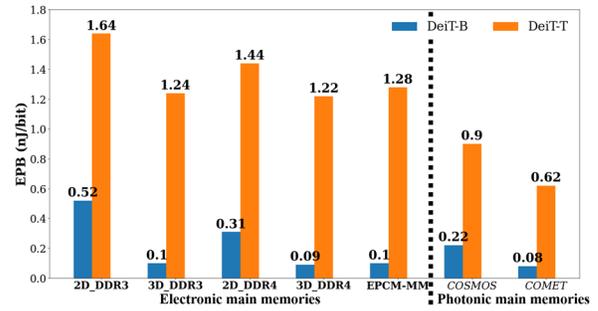

Fig. 10. EPB of DOTA accelerator with different main memories.

*COMET* is able to contribute towards better overall energy efficiency of systems it is part of, owing to the combination of high bandwidth and comparable EPB performance to other main memory platforms. We showcase this using a BW/EPB metric (see Fig. 9(c)), where *COMET* achieves 6.5× and 65.8× better BW/EPB over 3D_DDR4 (best electronic platform) and *COSMOS*, respectively.

### D. Photonic AI Accelerator Case Study

Photonic main memory architectures, such as our proposed *COMET*, are an excellent fit for emerging optical computing platforms. Over the recent years there has been several photonic computing platforms discussed [44]-[46]. However, to quantify the benefits of *COMET*, we consider DOTA [47], a photonic tensor engine-based transformer accelerator [48]. We analyzed how different electronic and photonic main memories impact the operation of. For this analysis, we considered the two transformer models DeiT-T and DeiT-B, as used in [47]. Fig. 10 shows the EPB results for the various main memory architectures considered. Photonic memory architectures not only provide higher bandwidth (as discussed earlier and shown in Fig. 9) than electronic memories, but also have the added benefit of being able to inject data directly into the photonic tensor engine, without the need for a energy-hungry electro-photonic conversion stages. *COMET*+DOTA achieves 1.3× and 2.06× lower EPB against 3D_DDR4+DOTA and 2.7× and 1.45× better EPB against *COSMOS*+DOTA. These results highlight the promise of photonic main memory for improving the performance of emerging optical computing platforms.

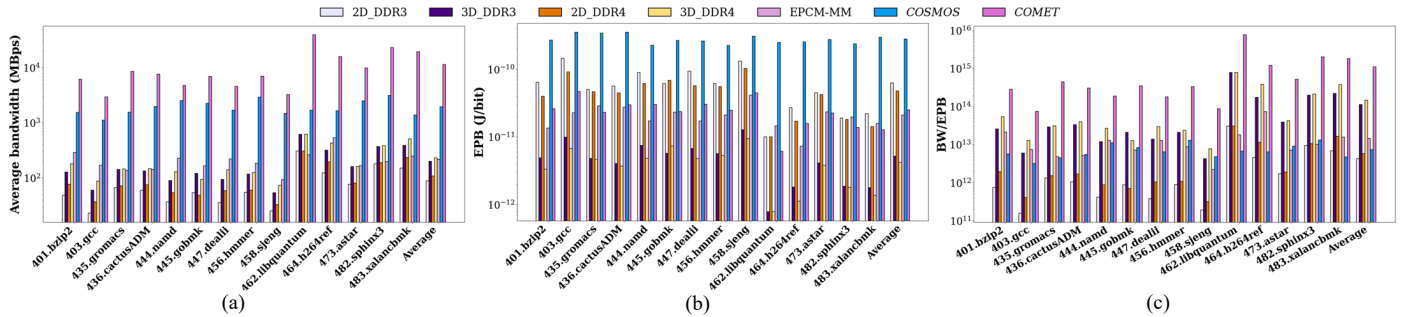

Fig. 9. (a) Average bandwidth (BW); (b) energy-per-bit (EPB); and (c) BW/EPB, of applications across memory architectures.

## V. CONCLUSIONS

In this work, we presented *COMET*, a low-loss, low-latency, and high throughput OPCM-based main memory architecture that makes use of GST integrated with silicon-on-insulator strip waveguides. We described the cross-layer design and optimization of our *COMET* architecture from the material and device level to the architecture level. Our GST cell was designed and optimized by leveraging transient unsteady state heat transfer equations integrated with finite-difference time domain simulations. By using silicon, unlike silicon nitride platforms, our GST cell offered a high transmission contrast between crystalline and amorphous state of the cell (≈96%). In addition, the designed cell offers 16 distinctive transmission levels with 6% spacing which makes *COMET* tolerant to transmission drift. Crosstalk-free, reliable memory operation was enabled with various loss-aware optimizations at the architecture level. Owing to these optimizations, *COMET* consumes only 26% of the power when compared to the best-known prior work on OPCM-based main memory architecture design. The low power consumption along with high-speed GST cell operations enable *COMET* to offer 7.1× better bandwidth, 15.5× lower EPB, and 3× lower latencies than the state-of-the-art OPCM-based main memory architecture. Our ongoing work is exploring the integration of approaches to reduce optical crosstalk [49]-[51] in the proposed OPCM-based architecture.